\documentclass{jpsj3}

\title{Coherence effect in a two-band superconductor: Application to iron pnictides}

\author{Keisuke Masuda\thanks{E-mail: masuda@kh.phys.waseda.ac.jp} and Susumu Kurihara 
}
\inst{Department of Physics, Waseda University, Okubo, Shinjuku, Tokyo 169-8555, Japan 
}
\abst{From a theoretical point of view, we propose an experimental method to determine the pairing symmetry of iron pnictides.
We focus on two kinds of pairing symmetries, $s_{+-}$ and $s_{++}$, which are strong candidates for the pairing symmetry of iron pnictides.
For each of these two symmetries, we calculate both the density and spin response functions by using the two-band BCS model within the one-loop approximation.
As a result, a clear difference is found between the $s_{+-}$- and $s_{++}$-wave states in the temperature dependence of the response functions at nesting vector $\bf{Q}$, which connects the hole and electron Fermi surfaces.
We point out that this difference comes from the coherence effect in the two-band superconductor.
We suggest that the pairing symmetry could be clarified by observing the temperature dependence of both the density and spin structure factors at the nesting vector $\bf{Q}$ in neutron scattering measurements.}

\kword{iron-pnictide superconductors, coherence effect, pairing symmetry, response functions, two-band BCS model}

\begin{document}
\maketitle

\section{\label{1}Introduction}

Recently discovered iron pnictides\cite{Kamihara} are expected to be unconventional (non-phonon-mediated) superconductors.
For such superconductors, determining the pairing symmetry is of great importance, because the pairing symmetry is closely related to the mechanism of superconductivity.

In iron pnictides, many theoretical studies have focused on the pairing symmetry.
In the early stage, Mazin {\it et al.}\cite{Mazin} and Kuroki {\it et al.}\cite{Kuroki} suggested that the $s_{+-}$-wave state, in which the $s$-wave gap on the hole Fermi surface has the opposite sign as that on the electron Fermi surface, is favored due to the nesting between the hole and electron Fermi surfaces.
This $s_{+-}$-wave state has also been supported by other studies; the theory based on the random-phase approximation (RPA) using the 16-band $d$-$p$ model\cite{Yanagi}, the perturbation theory\cite{Nomura}, and the functional renormalization-group theory\cite{Chubukov, Wang}. On the other hand, within the fluctuation exchange approximation (FLEX), it has been shown that the $s_{+-}$-wave state is not necessarily favored\cite{Arita}.
Onari and Kontani pointed out that the $s_{+-}$-wave state is quite fragile against nonmagnetic impurities, and suggested that the $s_{++}$-wave state, in which the $s$-wave gap on the hole Fermi surface has the same sign as that on the electron Fermi surface, is a promising candidate for iron pnictides\cite{Onari}.
Recently, by the microscopic calculations, they showed that the $s_{++}$-wave state is induced by orbital fluctuations due to the electron-phonon interaction\cite{Kontani}.

Many experiments have also been performed to clarify the character of iron pnictides.
The disconnected Fermi surfaces (some hole and some electron Fermi surfaces) have been observed by angle-resolved photoemission spectroscopy (ARPES).\cite{Liu,Sato}
In addition, the fully gapped $s$-wave nature has been confirmed by ARPES\cite{Ding,Nakayama,Terashima} and penetration depth measurements\cite{Hashimoto}.
However, there are few experimental methods which identify the phase difference between the superconducting gap on the hole Fermi surface and that on the electron Fermi surface.
Therefore, it is essential to suggest an experimental method to reveal the relative phase between the two superconducting gaps.

In this paper, we suggest an experimental method which distinguishes between the $s_{+-}$- and $s_{++}$-wave states.
In order to find the difference between these two states, we focus on the coherence effect in the response functions.
The coherence effect is one of the most important evidences that determined the success of the BCS theory.\cite{Schrieffer}
Experimentally, a qualitative difference between the behavior of the acoustic attenuation rate\cite{Bommel}
and that of the nuclear spin relaxation rate\cite{Hebel} had been established.
\begin{figure}
\begin{center}
\includegraphics{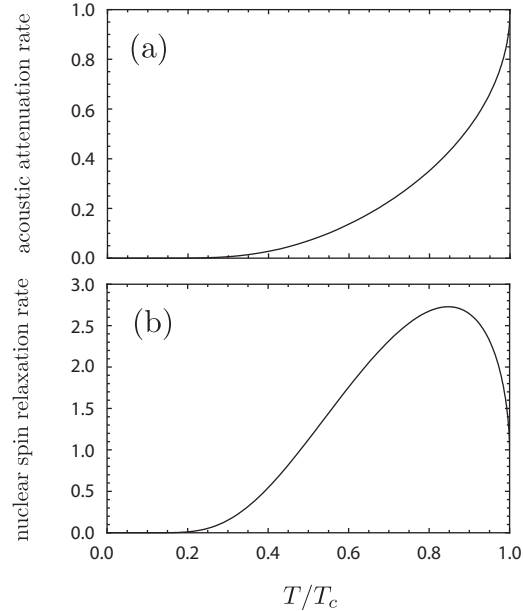}
\end{center}
\caption{(a) Schematic temperature dependence of the acoustic attenuation rate.
(b) Schematic temperature dependence of the nuclear spin relaxation rate.}
\label{fig1}
\end{figure}
The former decreases monotonically as temperature decreases [see Fig. \ref{fig1}(a)].
In contrast, the latter has a peak (called the ``coherence peak'') just below $T_{c}$ [see Fig. \ref{fig1}(b)].
Although the phenomenological two-fluid model could not explain both behaviors,\cite{Tinkham}
the BCS theory gave the correct explanation of both features.

According to the BCS theory, these behaviors can be understood as a result of the interference effect (called the ``coherence effect'') of the field-induced fluctuations in superconductors.
The temperature dependence of the response function is determined by the time reversal symmetry of the perturbing field.\cite{Cooper}
The perturbing field which relates to density fluctuations is even under time reversal.
As a result, the density response function which corresponds to the acoustic attenuation rate has the following coherence factor.
\begin{eqnarray}
\nonumber \frac{1}{2}\left( 1-\frac{{\Delta}_{\bf{k}}{\Delta}_{\bf{k+q}}}{E_{\bf{k}}E_{\bf{k+q}}} \right)
\end{eqnarray}
This coherence factor has a value close to 0 near the Fermi level, and cancels the square-root divergence in the density of states (DOS) which occurs in the superconducting state.
Hence, the density response function shows a monotonic decrease as temperature decreases.
In contrast, the perturbing field which relates to spin fluctuations is odd under time reversal.
As a consequence, the spin response function which corresponds to the nuclear spin relaxation rate includes the following coherence factor.
\begin{eqnarray}
\nonumber \frac{1}{2}\left( 1+\frac{{\Delta}_{\bf{k}}{\Delta}_{\bf{k+q}}}{E_{\bf{k}}E_{\bf{k+q}}} \right)
\end{eqnarray}
This coherence factor has a value close to 1 near the Fermi level, and does not cancel the divergence in the DOS.
Hence, the spin response function has a peak just below $T_{c}$.
This is the theoretical explanation of the coherence effect in the single-band $s$-wave superconductor.
It is worth noticing that the opposite behavior of the two different response functions can be explained within a framework of the BCS theory.
This is the reason why the coherence effect has become the strong evidence for the BCS theory.

To investigate the coherence effect in iron pnictides, let us extend the above analysis to multi-band cases.
First we introduce the two-band BCS model, and then calculate both the density and spin response functions within the one-loop approximation.
As mentioned above, the coherence factor includes the product of the two gap functions ($\Delta_{\bf{k}}\Delta_{\bf{k+q}}$).
In the $s_{+-}$-wave state, this product has a negative value ($\Delta_{\bf{k}}\Delta_{\bf{k+q}}<0$) when $\bf{q}$ equals $\bf{Q}$, which is the nesting vector connects the hole and electron Fermi surfaces.
Then the sign in front of the fraction in the coherence factor is {\it effectively reversed}.
Therefore, in this state, it is expected that the density (spin) response function has a peak just below $T_{c}$ (has a monotonic decrease as temperature decreases) as opposed to the single-band $s$-wave case.
In fact, these behaviors are also confirmed in our calculations.
On the other hand, in the $s_{++}$-wave state, the product of the two gap functions has a positive value ($\Delta_{\bf{k}}\Delta_{\bf{k+q}}>0$).
Therefore the sign in front of the fraction in the coherence factor does not change, and the behavior of the response functions is the same as in the single-band $s$-wave case.

From our analysis it follows that the pairing state would be identified by observing the temperature dependence of both the density and spin response functions (density and spin structure factors in actual experiments) at the nesting vector $\bf{Q}$.

\section{\label{2}Two-band BCS model}
In this section, we explain the two-band BCS model used in the calculation of the response functions.
As a starting point we use the tight binding model proposed by Raghu {\it et al.}\cite{Raghu}
This model includes two atomic orbitals ($d_{xz}$ and $d_{yz}$),
and the hopping integrals up to the second nearest neighbors.
The tight binding Hamiltonian $\mathcal{H}_{0}$ is given by
\begin{equation}
\!\mathcal{H}_{0}\!=\!\sum_{\mathbf{k}\sigma}\psi^{\dagger}_{\sigma}\!(\mathbf{k})\left[\left(\epsilon_{+}(\mathbf{k})\!-\!\mu\right){1}+\epsilon_{-}(\mathbf{k}){\tau}_{3}+\epsilon_{xy}(\mathbf{k}){\tau}_{1}\right]\psi_{\sigma}(\mathbf{k}),
\label{eq1}
\end{equation}
with
\begin{eqnarray}
\psi_{\sigma}(\mathbf{k}) = \left( \begin{array}{c}
                                   d_{x\sigma}(\mathbf{k}) \\
                                   d_{y\sigma}(\mathbf{k})
                                   \end{array} 
                            \right), 
\end{eqnarray}
where $d_{x\sigma}(\mathbf{k})$ ($d_{y\sigma}(\mathbf{k})$) is the annihilation operator with spin $\sigma$ and wave vector $\mathbf{k}$ in the orbital $d_{xz}$ ($d_{yz}$).
$\tau_{i}$ are the Pauli matrices, and
\begin{eqnarray}
\nonumber\epsilon_{\pm}(\mathbf{k})\!\!&=&\!\!\frac{1}{2}\left[\epsilon_{x}(\mathbf{k})\pm\epsilon_{y}(\mathbf{k})\right],\\
\nonumber\epsilon_{x}(\mathbf{k})\!\!&=&\!\!-2{t}_{1}\cos{k_{x}}-2{t}_{2}\cos{k_{y}}-4{t}_{3}\cos{k_{x}}\cos{k_{y}},\\
\nonumber\epsilon_{y}(\mathbf{k})\!\!&=&\!\!-2{t}_{2}\cos{k_{x}}-2{t}_{1}\cos{k_{y}}-4{t}_{3}\cos{k_{x}}\cos{k_{y}},\\
\epsilon_{xy}(\mathbf{k})\!\!&=&\!\!-4{t}_{4}\sin{k_{x}}\sin{k_{y}}.
\end{eqnarray}
Here, $t_{1}$ and $t_{2}$ ($t_{3}$ and $t_{4}$) are the nearest-neighbor (next-nearest-neighbor) hopping integrals.

We introduce the band operator $\gamma_{\nu\sigma}(\mathbf{k})$ which annihilates an electron with spin $\sigma$ and wave vector $\mathbf{k}$ in the band $\nu$:
\begin{equation}
d_{r\sigma}(\mathbf{k})=\sum_{\nu=\pm}a^{r}_{\nu}(\mathbf{k})\gamma_{\nu\sigma}(\mathbf{k}),
\end{equation}
with
\begin{eqnarray}
\nonumber a^{x}_{+}(\mathbf{k})&=&a^{y}_{-}(\mathbf{k})=\mathrm{sgn}\left[\epsilon_{xy}(\mathbf{k})\right]\sqrt{\frac{1}{2}+\frac{\epsilon_{-}(\mathbf{k})}{2\sqrt{\epsilon^{2}_{-}(\mathbf{k})+\epsilon^{2}_{xy}(\mathbf{k})}}},\\
a^{y}_{+}(\mathbf{k})&=&-a^{x}_{-}(\mathbf{k})=\sqrt{\frac{1}{2}-\frac{\epsilon_{-}(\mathbf{k})}{2\sqrt{\epsilon^{2}_{-}(\mathbf{k})+\epsilon^{2}_{xy}(\mathbf{k})}}}.
\end{eqnarray}
As a result, the tight binding Hamiltonian (\ref{eq1}) can be diagonalized:
\begin{equation}
\mathcal{H}_{0}=\sum_{\mathbf{k},\sigma,\nu=\pm}\xi_{\nu}(\mathbf{k})\gamma^{\dagger}_{\nu\sigma}(\mathbf{k})\gamma_{\nu\sigma}(\mathbf{k}),
\end{equation}
where the band energies are given by
\begin{equation}
\xi_{\pm}(\mathbf{k})=\epsilon_{+}(\mathbf{k})\pm\sqrt{\epsilon^{2}_{-}(\mathbf{k})+\epsilon^{2}_{xy}(\mathbf{k})}-\mu.
\end{equation}
\begin{figure}
\begin{center}
\includegraphics{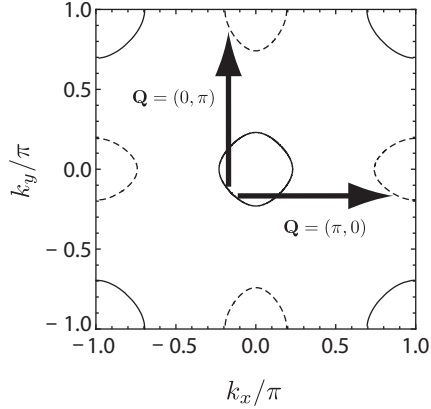}
\end{center}
\caption{The Fermi surface of the two-band model proposed by Raghu {\it et al.}
The solid (dashed) lines represent the hole (electron) Fermi surfaces, and the arrows indicate the nesting vectors.}
\label{fig2}
\end{figure}
In Fig. \ref{fig2}, we show the Fermi surface of this two-band model.
The vectors, $(0,\pi)$ and $(\pi,0)$, are the nesting vectors which connect the hole and electron Fermi surfaces.
Throughout this paper we use the parameters of Ref. [\citen{Raghu}]: $t_{1}=-1.0$, $t_{2}=1.3$, $t_{3}=t_{4}=-0.85$, and $\mu=1.45$.

By adding the BCS-type interaction Hamiltonian $\mathcal{H}'$, we obtain the following two-band BCS Hamiltonian $\mathcal{H}$:
\begin{eqnarray}
\!\!\mathcal{H}\!\!&=&\!\!\mathcal{H}_{0}+\mathcal{H}',\label{eq8}\\
\!\!\mathcal{H}'\!\!\!&=&\!\!\!-\!\!\!\!\sum_{\mathbf{k},\nu=\pm}\!\left[\Delta_{\nu}\gamma^{\dagger}_{\nu\uparrow}(\mathbf{k})\gamma^{\dagger}_{\nu\downarrow}(-\mathbf{k})\!+\!\Delta_{\nu}\gamma_{\nu\downarrow}(-\mathbf{k})\gamma_{\nu\uparrow}(\mathbf{k})\right],
\end{eqnarray}
where $\Delta_{\nu}$ is the superconducting gap of the band $\nu$.
In this model, the condition $\Delta_{+}=-\Delta_{-}$ ($\Delta_{+}=\Delta_{-}$) corresponds to the $s_{+-}$-wave state ($s_{++}$-wave state).
Using this Hamiltonian (\ref{eq8}), the normal and anomalous Green's functions in the band $\nu$ are as follows:
\begin{eqnarray}
\nonumber\mathcal{G}_{\nu}(\mathbf{k},i{\omega}_{n})\!\!&=&\!\!\frac{i{\omega}_{n}+\xi_{\nu}(\mathbf{k})}{{(i{\omega}_{n})}^{2}-E^{2}_{\nu}(\mathbf{k})},\\
\mathcal{F}_{\nu,\sigma\bar{\sigma}}(\mathbf{k},i{\omega}_{n})\!\!&=&\!\!{F}^{\dagger}_{\nu,\bar{\sigma}\sigma}(\mathbf{k},i{\omega}_{n})\!=\!\frac{{\sigma}{\Delta}_{\nu}}{{(i{\omega}_{n})}^{2}-E^{2}_{\nu}(\mathbf{k})},
\label{eq10}
\end{eqnarray}
where $E_{\nu}(\mathbf{k})=\sqrt{\xi^{2}_{\nu}(\mathbf{k})+{\Delta}^2_{\nu}}$ and $\sigma=\,\uparrow,\downarrow\,=\!1,-1$.
In numerical calculations, we use the following interpolation formula:
\begin{equation}
{\Delta}_{\nu}\simeq{\Delta}^{(0)}_{\nu}\tanh{\left(1.74\sqrt{\frac{{T}_{c}}{T}-1}\,\right)}.
\end{equation}

\section{\label{3}Calculation of the response functions}

Now we calculate the response functions for the two-band BCS model.
By using the Kubo formula, the density response function can be written as
\begin{equation}
D(\mathbf{q},i{\omega}_{m})=-\sum_{rt}\int^{\beta}_{0}d{\tau}e^{i{\omega}_{m}{\tau}}{\langle}T_{\tau}{\rho}_{r}(\mathbf{q},\tau){\rho}_{t}(\mathbf{-q},0){\rangle},
\label{eq12}
\end{equation}
where $r,t$ denote the orbital indices, and ${\rho}_{r}(\mathbf{q},\tau)$ is defined by
\begin{equation}
{\rho}_{r}(\mathbf{q},\tau)=\sum_{\mathbf{k}{\sigma}}d^{\dagger}_{r{\sigma}}(\mathbf{k+q},\tau)d_{r{\sigma}}(\mathbf{k},\tau).
\end{equation}
Rewriting Eq. (\ref{eq12}) in the band representation, we have
\begin{eqnarray}
\nonumber\!\!\!\!\!\!\!\!\!\!\!\!&D&\!\!\!(\mathbf{q},i{\omega}_{m})\,\!=\!-\sum_{\mathbf{kk'}}\sum_{{\sigma}{\sigma}'}\sum_{rt}\sum_{{\nu}_{1},\cdots,{\nu}_{4}}\int^{\beta}_{0}d{\tau}e^{i{\omega}_{m}{\tau}}\\
\nonumber&{\times}&\!\!a^{r}_{{\nu}_{1}}(\mathbf{k+q})a^{r}_{{\nu}_{2}}(\mathbf{k})a^{t}_{{\nu}_{3}}(\mathbf{k'-q})a^{t}_{{\nu}_{4}}(\mathbf{k'})\\
&{\times}&\!\!\!\!{\langle}T_{\tau}{\gamma}^{\dagger}_{{\nu}_{1}{\sigma}}(\mathbf{k+q},\tau){\gamma}_{{\nu}_{2}{\sigma}}(\mathbf{k},\tau){\gamma}^{\dagger}_{{\nu}_{3}{{\sigma}'}}(\mathbf{k'-q},0){\gamma}_{{\nu}_{4}{{\sigma}'}}(\mathbf{k'},0){\rangle}.
\end{eqnarray}

Now we consider the one-loop contribution to the density response function.
Then we obtain the following expression:
\begin{eqnarray}
\nonumber D(\mathbf{q},i{\omega}_{m})\!\!&=&\!\!\frac{1}{\beta}\sum_{n}\sum_{\mathbf{k}{\sigma}}\sum_{rt}\sum_{{\nu}{\nu}'}\\
\nonumber&\times&\!a^{r}_{\nu}(\mathbf{k+q})a^{r}_{{\nu}'}(\mathbf{k})a^{t}_{{\nu}'}(\mathbf{k})a^{t}_{\nu}(\mathbf{k+q})\\
\nonumber&\times&\![ \mathcal{G}_{{\nu}'}(\mathbf{k},i{\omega}_{n})\mathcal{G}_{\nu}(\mathbf{k+q},i{\omega}_{n}-i{\omega}_{m})\\
&-&\!\mathcal{F}_{{\nu}',\bar{{\sigma}}{\sigma}}(\mathbf{k},i{\omega}_{n})\mathcal{F}^{\dagger}_{{\nu},{\sigma}\bar{{\sigma}}}(\mathbf{k+q},i{\omega}_{n}+i{\omega}_{m}) ].
\label{eq15}
\end{eqnarray}
Here, we substitute the normal and anomalous Green's functions (\ref{eq10}) into Eq. (\ref{eq15}).
Summing over the Matsubara frequencies and performing the analytic continuation $i{\omega}_{m}\,{\rightarrow}\,\,{\omega}+i{\delta}$, we finally obtain
\begin{eqnarray}
\nonumber\!\!\!\!\!\!\!\!\!\!\!\!\!\!\!\!D(\mathbf{q},{\omega}+i{\delta})\!\!\!\!\!\!&=&\!\!\!\!\!\!\sum_{\mathbf{k}}\sum_{rt}\sum_{{\nu}{\nu}'}a^{r}_{\nu}(\mathbf{k+q})a^{r}_{{\nu}'}(\mathbf{k})a^{t}_{{\nu}'}(\mathbf{k})a^{t}_{\nu}(\mathbf{k+q})\\
\nonumber\times&&\!\!\!\!\!\!\!\!\!\!\!\!\left[\frac{1-f\left(E_{\nu'}(\mathbf{k})\right)-f\left(E_{\nu}(\mathbf{k+q})\right)}{\omega-E_{{\nu}'}(\mathbf{k})-E_{\nu}(\mathbf{k+q})+i{\delta}}\frac{1}{2}\left\{\left(1+\frac{\xi_{{\nu}'}(\mathbf{k})}{E_{{\nu}'}(\mathbf{k})}\right)\left(1-\frac{\xi_{\nu}(\mathbf{k+q})}{E_{\nu}(\mathbf{k+q})}\right)+\frac{{\Delta}_{{\nu}'}{\Delta}_{\nu}}{E_{{\nu}'}(\mathbf{k})E_{\nu}(\mathbf{k+q})}\right\} \right.\\
\nonumber&-&\!\!\frac{1-f\left(E_{\nu'}(\mathbf{k})\right)-f\left(E_{\nu}(\mathbf{k+q})\right)}{\omega+E_{{\nu}'}(\mathbf{k})+E_{\nu}(\mathbf{k+q})+i{\delta}}\frac{1}{2}\left\{\left(1-\frac{\xi_{{\nu}'}(\mathbf{k})}{E_{{\nu}'}(\mathbf{k})}\right)\left(1+\frac{\xi_{\nu}(\mathbf{k+q})}{E_{\nu}(\mathbf{k+q})}\right)+\frac{{\Delta}_{{\nu}'}{\Delta}_{\nu}}{E_{{\nu}'}(\mathbf{k})E_{\nu}(\mathbf{k+q})}\right\}\\
\nonumber&-&\!\!\frac{f\left(E_{\nu'}(\mathbf{k})\right)-f\left(E_{\nu}(\mathbf{k+q})\right)}{\omega-E_{{\nu}'}(\mathbf{k})+E_{\nu}(\mathbf{k+q})+i{\delta}}\frac{1}{2}\left\{\left(1+\frac{\xi_{{\nu}'}(\mathbf{k})}{E_{{\nu}'}(\mathbf{k})}\right)\left(1+\frac{\xi_{\nu}(\mathbf{k+q})}{E_{\nu}(\mathbf{k+q})}\right)-\frac{{\Delta}_{{\nu}'}{\Delta}_{\nu}}{E_{{\nu}'}(\mathbf{k})E_{\nu}(\mathbf{k+q})}\right\}\\
&+&\!\!\left.\frac{f\left(E_{\nu'}(\mathbf{k})\right)-f\left(E_{\nu}(\mathbf{k+q})\right)}{\omega+E_{{\nu}'}(\mathbf{k})-E_{\nu}(\mathbf{k+q})+i{\delta}}\frac{1}{2}\left\{\left(1-\frac{\xi_{{\nu}'}(\mathbf{k})}{E_{{\nu}'}(\mathbf{k})}\right)\left(1-\frac{\xi_{\nu}(\mathbf{k+q})}{E_{\nu}(\mathbf{k+q})}\right)-\frac{{\Delta}_{{\nu}'}{\Delta}_{\nu}}{E_{{\nu}'}(\mathbf{k})E_{\nu}(\mathbf{k+q})}\right\}\right]\!,~~~~~~
\label{eq16}
\end{eqnarray}
where $f(E)=1/(e^{{\beta}E}+1)$ is the Fermi distribution function.
The first and second terms in the bracket of Eq. (\ref{eq16}) are due to the pair creation and annihilation.
On the other hand, the third and fourth terms are due to the scattering of thermally excited quasiparticles.
The factor $\frac{1}{2}\{\cdots\}$ is the so-called coherence factor.
Four coherence factors are included in Eq. (\ref{eq16}).

Next we calculate the spin response function.
The spin response function is given by 
\begin{equation}
\!\!\!{\chi}^{+-}(\mathbf{q},i{\omega}_{m})\!=\!-\!\sum_{rt}\int^{\beta}_{0}\!\!\!d{\tau}e^{i{\omega}_{m}{\tau}}{\langle}T_{\tau}S^{+}_{r}(\mathbf{q},\tau)S^{-}_{t}(\mathbf{-q},0){\rangle},
\end{equation}
where $r,t$ also denote the orbital indices.
Here, $S^{+}_{r}(\mathbf{q},\tau)$ and $S^{-}_{r}(\mathbf{q},\tau)$ are defined by
\begin{equation}
S^{+}_{r}(\mathbf{q},\tau)=\left( S^{-}_{r}(-\mathbf{q},\tau) \right)^{\dagger}=\sum_{\mathbf{k}}d^{\dagger}_{r{\uparrow}}(\mathbf{k+q},\tau)d_{r{\downarrow}}(\mathbf{k},\tau).
\end{equation}
The spin response function can be calculated
in the same way as the density response function.
The final expression for the spin response function is as follows. 
\begin{eqnarray}
\nonumber\!\!\!\!\!\!\!\!\!\!\!\!\!\!\!\!{\chi}^{+-}(\mathbf{q},{\omega}+i{\delta})\!\!\!\!\!\!&=&\!\!\!\!\!\!\frac{1}{2}\sum_{\mathbf{k}}\sum_{rt}\sum_{{\nu}{\nu}'}a^{r}_{\nu}(\mathbf{k+q})a^{r}_{{\nu}'}(\mathbf{k})a^{t}_{{\nu}'}(\mathbf{k})a^{t}_{\nu}(\mathbf{k+q})\\
\nonumber\times&&\left[\frac{1-f\left(E_{\nu'}(\mathbf{k})\right)-f\left(E_{\nu}(\mathbf{k+q})\right)}{\omega-E_{{\nu}'}(\mathbf{k})-E_{\nu}(\mathbf{k+q})+i{\delta}}\frac{1}{2}\left\{\left(1+\frac{\xi_{{\nu}'}(\mathbf{k})}{E_{{\nu}'}(\mathbf{k})}\right)\left(1-\frac{\xi_{\nu}(\mathbf{k+q})}{E_{\nu}(\mathbf{k+q})}\right)-\frac{{\Delta}_{{\nu}'}{\Delta}_{\nu}}{E_{{\nu}'}(\mathbf{k})E_{\nu}(\mathbf{k+q})}\right\} \right. \\
\nonumber&-&\!\!\frac{1-f\left(E_{\nu'}(\mathbf{k})\right)-f\left(E_{\nu}(\mathbf{k+q})\right)}{\omega+E_{{\nu}'}(\mathbf{k})+E_{\nu}(\mathbf{k+q})+i{\delta}}\frac{1}{2}\left\{\left(1-\frac{\xi_{{\nu}'}(\mathbf{k})}{E_{{\nu}'}(\mathbf{k})}\right)\left(1+\frac{\xi_{\nu}(\mathbf{k+q})}{E_{\nu}(\mathbf{k+q})}\right)-\frac{{\Delta}_{{\nu}'}{\Delta}_{\nu}}{E_{{\nu}'}(\mathbf{k})E_{\nu}(\mathbf{k+q})}\right\}\\
\nonumber&-&\!\!\frac{f\left(E_{\nu'}(\mathbf{k})\right)-f\left(E_{\nu}(\mathbf{k+q})\right)}{\omega-E_{{\nu}'}(\mathbf{k})+E_{\nu}(\mathbf{k+q})+i{\delta}}\frac{1}{2}\left\{\left(1+\frac{\xi_{{\nu}'}(\mathbf{k})}{E_{{\nu}'}(\mathbf{k})}\right)\left(1+\frac{\xi_{\nu}(\mathbf{k+q})}{E_{\nu}(\mathbf{k+q})}\right)+\frac{{\Delta}_{{\nu}'}{\Delta}_{\nu}}{E_{{\nu}'}(\mathbf{k})E_{\nu}(\mathbf{k+q})}\right\}\\                                               
&+& \left. \!\!\frac{f\left(E_{\nu'}(\mathbf{k})\right)-f\left(E_{\nu}(\mathbf{k+q})\right)}{\omega+E_{{\nu}'}(\mathbf{k})-E_{\nu}(\mathbf{k+q})+i{\delta}}\frac{1}{2}\left\{\left(1-\frac{\xi_{{\nu}'}(\mathbf{k})}{E_{{\nu}'}(\mathbf{k})}\right)\left(1-\frac{\xi_{\nu}(\mathbf{k+q})}{E_{\nu}(\mathbf{k+q})}\right)+\frac{{\Delta}_{{\nu}'}{\Delta}_{\nu}}{E_{{\nu}'}(\mathbf{k})E_{\nu}(\mathbf{k+q})}\right\}\right]\!.~~~~~~
\label{eq19}
\end{eqnarray}

Now we present an explanation for the coherence factors.
The following two regions are considered here:
(i) $\omega\approx|\Delta^{(0)}_{\nu}|+|\Delta^{(0)}_{\nu'}|$ and $T{\ll}T_{c}$.
(ii) $\omega{\ll}|\Delta^{(0)}_{\nu}|+|\Delta^{(0)}_{\nu'}|$ and $0{\leq}T{\leq}T_{c}$.
In the region (i), the dominant terms in the bracket of each response function are the first and second ones (i.e., pair creation and annihilation terms).
This is because the third and fourth terms vanish due to the fact that $f\left(E_{\nu'}(\mathbf{k})\right)-f\left(E_{\nu}(\mathbf{k+q})\right)\approx0$ for $T{\ll}T_{c}$.
Hence we focus on the following coherence factor:
\begin{eqnarray}
\!\!\!\!\!\frac{1}{2}\!\left\{\!\!\left(\!1+\frac{\xi_{{\nu}'}(\mathbf{k})}{E_{{\nu}'}(\mathbf{k})}\!\right)\!\!\!\left(\!1-\frac{\xi_{\nu}(\mathbf{k+q})}{E_{\nu}(\mathbf{k+q})}\!\right)\!{\pm}\frac{{\Delta}_{{\nu}'}{\Delta}_{\nu}}{E_{{\nu}'}(\mathbf{k})E_{\nu}(\mathbf{k+q})}\!\!\right\}\!\!.
\label{eq20}
\end{eqnarray}
The upper (lower) sign in (\ref{eq20}) corresponds to the case of the density (spin) response function.
In the $s_{+-}$-wave state ($\Delta_{\nu'}\Delta_{\nu}<0$), the sign in front of the term $\Delta_{\nu'}\Delta_{\nu}/E_{\nu'}E_{\nu}$ in the coherence factor at $\bf{q}=\bf{Q}$ (nesting vector) is effectively reversed.
As already pointed out by Korshunov {\it et al.}\cite{Korshunov} and Maier {\it et al,}\cite{Maier1,Maier2} this sign reversal causes the resonance peak in inelastic neutron scattering measurements\cite{remark1}. 
In contrast, the same does not hold for the $s_{++}$-wave state.
The experiment by Christianson {\it et al.}\cite{Christianson}  confirms the existence of such a resonance peak and is the positive information about the $s_{+-}$-wave state (see also \ref{5}. Discussion).

Next, we consider the region (ii) on which we particularly focus in this paper.
In this region, the main terms in the bracket of each response function are the third and fourth ones (i.e., quasiparticle scattering terms) because the delta functions $\delta\left(\omega-E_{{\nu}'}(\mathbf{k})-E_{\nu}(\mathbf{k+q})\right)$ and $\delta\left(\omega+E_{{\nu}'}(\mathbf{k})+E_{\nu}(\mathbf{k+q})\right)$
in the first and second terms are zero for almost all temperatures.
Therefore, we pay attention to the following coherence factor:
\begin{eqnarray}
\!\!\!\!\!\frac{1}{2}\!\left\{\!\!\left(\!1+\frac{\xi_{{\nu}'}(\mathbf{k})}{E_{{\nu}'}(\mathbf{k})}\!\right)\!\!\!\left(\!1+\frac{\xi_{\nu}(\mathbf{k+q})}{E_{\nu}(\mathbf{k+q})}\!\right)\!{\mp}\frac{{\Delta}_{{\nu}'}{\Delta}_{\nu}}{E_{{\nu}'}(\mathbf{k})E_{\nu}(\mathbf{k+q})}\!\!\right\}\!\!.
\label{eq21}
\end{eqnarray}
The upper (lower) sign in (\ref{eq21}) corresponds to the case of the density (spin) response function.
Note that in both the density and spin responses, the sign in front of the term $\Delta_{\nu'}\Delta_{\nu}/E_{\nu'}E_{\nu}$ in the coherence factor in the region (ii) is opposite to that in the region (i).
In the region (ii), such a sign of the coherence factor directly affects the temperature dependence of the response functions.
For example, we consider the case $\Delta_{\nu'}\Delta_{\nu}>0$ (the $s_{++}$-wave state).
In the density response, the coherence factor cancels the divergence in the DOS, and the corresponding response function (i.e., the density response function) decreases monotonically as temperature decreases.
In contrast, in the spin response, the coherence factor does not cancel the divergence in the DOS, and the corresponding response function (i.e., the spin response function) has a peak just below $T_{c}$.
In this sense, the coherence factor behaves like a ``switch'', which controls the contribution from the divergence in the DOS. 
This is the essential point of the coherence effect.

\begin{figure}
\begin{center}
\includegraphics{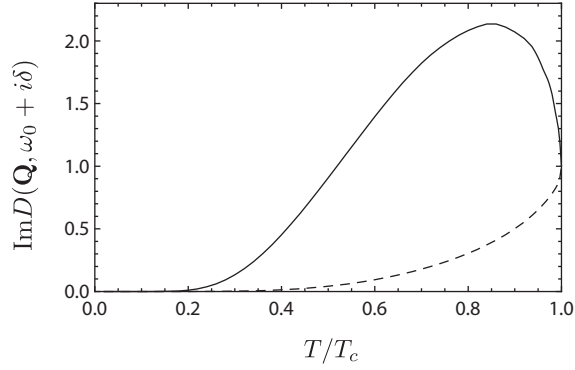}
\end{center}
\caption{Temperature dependence of the density response function normalized by its value at $T=T_{c}$ with $|{\Delta}^{(0)}_{+}|=|{\Delta}^{(0)}_{-}|=0.05$, ${T}_{c}=0.03$, ${\omega}_{0}=0.001$, $\mathbf{Q}=(\pi,0)$, and $\delta=0.0001$.
The solid (dashed) line represents the case of the ${s}_{+-}$-wave state (${s}_{++}$-wave state).}
\label{fig3}
\end{figure}
\begin{figure}
\begin{center}
\includegraphics{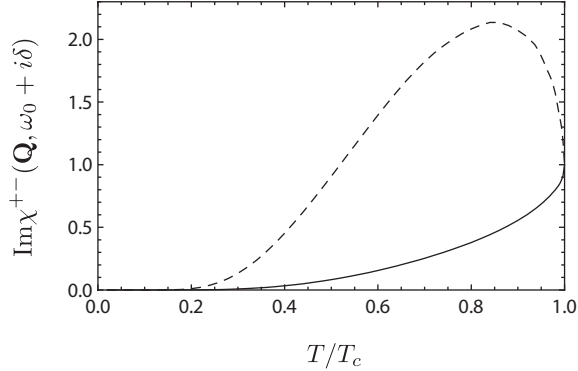}
\end{center}
\caption{Temperature dependence of the spin response function normalized by its value at $T=T_{c}$ with $|{\Delta}^{(0)}_{+}|=|{\Delta}^{(0)}_{-}|=0.05$, ${T}_{c}=0.03$, ${\omega}_{0}=0.001$, $\mathbf{Q}=(\pi,0)$, and $\delta=0.0001$.
The solid (dashed) line represents the case of the ${s}_{+-}$-wave state (${s}_{++}$-wave state).}
\label{fig4}
\end{figure}

In Fig. \ref{fig3}, we show the temperature dependence of the density response function.
In the $s_{+-}$-wave state, the density response has a peak just below $T_{c}$.
In contrast, in the $s_{++}$-wave state, the density response decreases monotonically as temperature decreases. 
In the $s_{+-}$-wave state, the gap of the band $\nu$ and that of the band $\nu'$ have the opposite sign because the nesting vector $\mathbf{Q}\big{(}\!\!=\!(\pi,0)\big{)}$ connects the hole and electron Fermi surfaces.
Therefore, the product of the gaps $\Delta_{\nu}\Delta_{\nu'}$ has a negative sign, and the upper sign of the coherence factor (\ref{eq21}) is effectively reversed.
As a result, the coherence factor does not cancel the divergence in the DOS, and the peak appears.
On the other hand, in the $s_{++}$-wave state, the gap of the band $\nu$ and that of the band $\nu'$ have the same sign.
Therefore, the product of the gaps has a positive sign,
and the upper sign of the coherence factor (\ref{eq21}) does not change.
As a consequence, the coherence factor cancels the divergence in the DOS, and the peak does not appear.

In Fig. \ref{fig4}, we show the temperature dependence of the spin response function.
In the $s_{+-}$-wave state, the spin response shows a monotonic decrease as temperature decreases.
In contrast, in the $s_{++}$-wave state, the spin response has a peak just below $T_{c}$.
In the $s_{+-}$-wave state, the lower sign of the coherence factor (\ref{eq21}) is effectively reversed, and the peak does not appear.
On the other hand, in the $s_{++}$-wave state, the lower sign of the coherence factor (\ref{eq21}) does not change, and the peak appears.

\begin{table}
\caption{The dominant coherence factors in the region (ii) depending on both the time reversal symmetry of the external field and the pairing state. We show the case of $\mathbf{q}=\mathbf{Q}=(\pi,0)$.}
\label{ta1}
\begin{center}
\footnotesize
\begin{tabular}{ccc}
\hline
& even parity & odd parity \\
& (density response function) & (spin response function) \\
\hline
$s_{+-}$-wave state &$\frac{1}{2}\left\{\!\left(1+\frac{\xi_{{\nu}'}(\mathbf{k})}{E_{{\nu}'}(\mathbf{k})}\right)\!\left(1+\frac{\xi_{\nu}(\mathbf{k+Q})}{E_{\nu}(\mathbf{k+Q})}\right)+\frac{|{\Delta}_{{\nu}'}||{\Delta}_{\nu}|}{E_{{\nu}'}(\mathbf{k})E_{\nu}(\mathbf{k+Q})}\!\right\}$ &$\frac{1}{2}\left\{\!\left(1+\frac{\xi_{{\nu}'}(\mathbf{k})}{E_{{\nu}'}(\mathbf{k})}\right)\!\left(1+\frac{\xi_{\nu}(\mathbf{k+Q})}{E_{\nu}(\mathbf{k+Q})}\right)-\frac{|{\Delta}_{{\nu}'}||{\Delta}_{\nu}|}{E_{{\nu}'}(\mathbf{k})E_{\nu}(\mathbf{k+Q})}\!\right\}$\\
$s_{++}$-wave state &$\frac{1}{2}\left\{\!\left(1+\frac{\xi_{{\nu}'}(\mathbf{k})}{E_{{\nu}'}(\mathbf{k})}\right)\!\left(1+\frac{\xi_{\nu}(\mathbf{k+Q})}{E_{\nu}(\mathbf{k+Q})}\right)-\frac{|{\Delta}_{{\nu}'}||{\Delta}_{\nu}|}{E_{{\nu}'}(\mathbf{k})E_{\nu}(\mathbf{k+Q})}\!\right\}$ &$\frac{1}{2}\left\{\!\left(1+\frac{\xi_{{\nu}'}(\mathbf{k})}{E_{{\nu}'}(\mathbf{k})}\right)\!\left(1+\frac{\xi_{\nu}(\mathbf{k+Q})}{E_{\nu}(\mathbf{k+Q})}\right)+\frac{|{\Delta}_{{\nu}'}||{\Delta}_{\nu}|}{E_{{\nu}'}(\mathbf{k})E_{\nu}(\mathbf{k+Q})}\!\right\}$\\
\hline
\end{tabular}
\end{center}
\end{table}
Here, we give a brief summary of the coherence effect in the two-band superconductor.
As mentioned above, it is the sign of the coherence factor that plays a crucial role in the coherence effect.
In the single-band s-wave superconductor, the sign of the coherence factor is determined only by the time reversal symmetry of the external field.
However, in the two-band superconductor, the effective sign of the coherence factor is determined by both the time reversal symmetry of the external field and the symmetry of the superconducting gap (See Table \ref{ta1}).
The coherence factor with the effective plus sign gives the coherence peak.
On the other hand, the coherence factor with the effective minus sign does not give the coherence peak.
It is worthy to note that in both pairing states the temperature dependence of the density response function is opposite from that of the spin response function.
As discussed in the next section, this point is quite important for the determination of the pairing symmetry.

For simplicity, we used the two-band BCS model in our calculation. However, this model cannot reproduce the detailed band structure of iron pnictides. This simple two-band model and the more realistic five-band model\cite{Kuroki} can lead to qualitatively different results. For example, in the analysis of the superconducting order parameters, Nomura demonstrated that the $s_{+-}$-wave state is favored for the five-band model, whereas the $p$-wave state is for the two-band model\cite{Nomura}. However, as we mentioned above, the behavior of the density and spin response functions in the two-band $s$-wave superconductor comes from both the peak in the DOS and the coherence factor, and it can be simply understood by an extension of the BCS theory for the single-band $s$-wave superconductor. As long as the superconducting order parameter in each band has a fully-gapped $s$-wave symmetry, the details of the band structure does not affect the qualitative behavior of both the response functions. Therefore, the qualitatively same results are also expected in the five-band BCS model.

\section{\label{4}Suggested experiment}

\begin{figure}
\begin{center}
\includegraphics{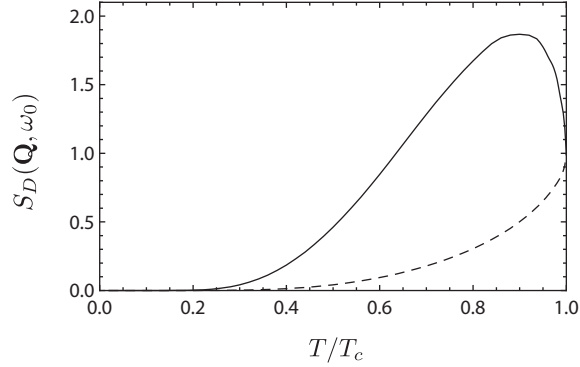}
\end{center}
\caption{Temperature dependence of the density structure factor normalized by its value at $T=T_{c}$ with $|{\Delta}^{(0)}_{+}|=|{\Delta}^{(0)}_{-}|=0.05$, ${T}_{c}=0.03$, ${\omega}_{0}=0.001$, $\mathbf{Q}=(\pi,0)$, and $\delta=0.0001$.
The solid (dashed) line represents the case of the ${s}_{+-}$-wave state (${s}_{++}$-wave state).}
\label{fig5}
\end{figure}
\begin{figure}
\begin{center}
\includegraphics{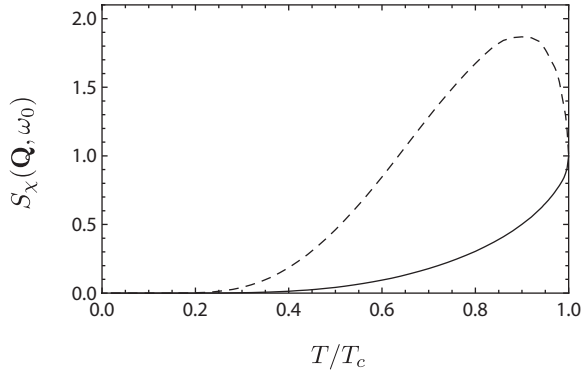}
\end{center}
\caption{Temperature dependence of the spin structure factor normalized by its value at $T=T_{c}$ with $|{\Delta}^{(0)}_{+}|=|{\Delta}^{(0)}_{-}|=0.05$, ${T}_{c}=0.03$, ${\omega}_{0}=0.001$, $\mathbf{Q}=(\pi,0)$, and $\delta=0.0001$.
The solid (dashed) line represents the case of the ${s}_{+-}$-wave state (${s}_{++}$-wave state).}
\label{fig6}
\end{figure}
In this section, we would like to present an experimental method which determines the correct symmetry of the order parameter; the $s_{+-}$- or $s_{++}$- wave state.
To accomplish this, we have to connect the theoretical calculations to the experimental observables.
The scattering cross section associated with the density (spin) response is proportional to the density (spin) structure factor.
The density structure factor $S_{D}(\mathbf{q},\omega)$ and the spin structure factor $S_{\chi}(\mathbf{q},\omega)$ are given by
\begin{eqnarray}
S_{D}(\mathbf{q},\omega)&=&-\frac{1}{\pi}\frac{1}{1-e^{-\beta\omega}}\textrm{Im}D(\mathbf{q},\omega+i\delta),\\
S_{\chi}(\mathbf{q},\omega)&=&-\frac{1}{\pi}\frac{1}{1-e^{-\beta\omega}}\textrm{Im}{\chi}^{+-}(\mathbf{q},\omega+i\delta).
\end{eqnarray}
Here, $D(\mathbf{q},\omega+i\delta)$ and ${\chi}^{+-}(\mathbf{q},\omega+i\delta)$ are the density and spin response functions, respectively.
The temperature dependence of the density (spin) structure factor is presented in Fig. \ref{fig5} (Fig. \ref{fig6}).
The behavior of the density (spin) structure factor is qualitatively the same as that of the density (spin) response function.

To distinguish between the $s_{+-}$- and $s_{++}$-wave states, it is necessary to measure the temperature dependence of both the density and spin structure factors at the nesting vector $\bf{Q}$.
If the following two behaviors are observed simultaneously, it is expected that the $s_{+-}$-wave state is realized:
(i) the density structure factor has a peak just below $T_{c}$. 
(ii) the spin structure factor decreases monotonically as temperature decreases.
On the other hand, if the following two behaviors are confirmed simultaneously, it is expected that the $s_{++}$-wave state is realized:
(i) the density structure factor decreases monotonically as temperature decreases.
(ii) the spin structure factor has a peak just below $T_{c}$. 

\begin{figure}
\begin{center}
\includegraphics{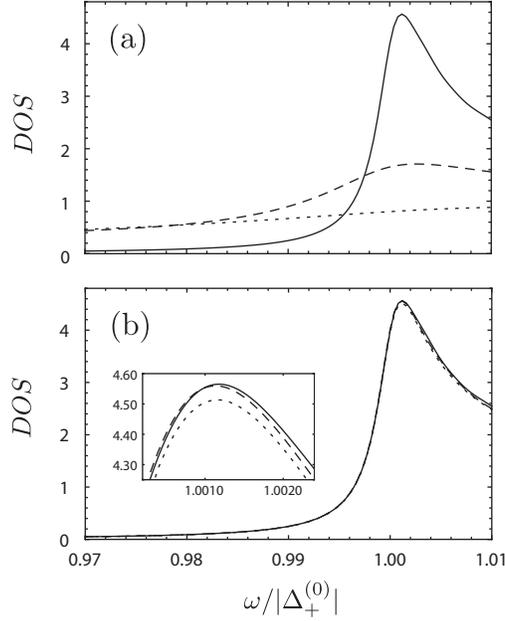}
\end{center}
\caption{The DOS in the (a) $s_{+-}$- and (b) $s_{++}$-wave states at $T=0$ for $n_{imp}=0.01$, $|{\Delta}^{(0)}_{+}|=|{\Delta}^{(0)}_{-}|=0.05$, ${T}_{c}=0.03$, and $\delta=0.0001$.
$I=0$ (solid line), $I=0.5$ (dashed line), and $I=4.5$ (dotted line). $1024\times1024$ $\bf{k}$-meshes are used in the numerical calculations.}
\label{fig7}
\end{figure}

In real experiments, however, there is no guarantee that the above-mentioned opposite behavior of the density and spin structure factors is always obtained.
This is because we cannot neglect the effect of impurities in iron pnictides.
Here we consider the influence of impurities on the coherence effect.
By using the T-matrix approximation, we calculate the effect of orbital-diagonal local impurities\cite{Onari} on the DOS of our two-band BCS model.
In Fig. \ref{fig7}(a) [Fig. \ref{fig7}(b)], we present the DOS in the $s_{+-}$-wave ($s_{++}$-wave) state.
In the $s_{+-}$-wave state, realistic impurity potentials eliminate the peak in the DOS, and makes the bound state inside the gap.
Meanwhile, in the $s_{++}$-wave state, the peak in the DOS is quite robust against impurities, and the in-gap state hardly appears.
If impurities eliminate the peak in the DOS, the opposite behavior of the density and spin structure factors is not observed, and both the structure factors decrease monotonically as temperature decreases.

As mentioned in the introduction, the acoustic attenuation and nuclear spin relaxation rates represent the density and spin responses, respectively.
However, the acoustic attenuation measurements observe the density response at small wave vector $\mathbf{q}\sim0$, and the nuclear spin relaxation measurements observe the spin response integrated over $\bf{q}$ in the Brillouin zone.
Namely, these two measurements are not suitable for measuring the density and spin responses at a large wave vector.
Therefore, we hope that the density and spin structure factors at the nesting vector $\bf{Q}$ would be observed by the inelastic neutron scattering measurements\cite{remark2}.
In the near future, if the resolution is improved, the density structure factor at the nesting vector $\bf{Q}$ may be observed by the inelastic X-ray scattering measurements.

At the end of this section we discuss the nuclear spin relaxation rate in iron pnictides.
In iron pnictides, the coherence peak in the nuclear spin relaxation rate has not been observed.\cite{Grafe,Matano,Nakai,Mukuda,Terasaki}
In the $s_{++}$-wave state, the coherence factor in the spin response at all wave vectors in the Brillouin zone does not cancel the peak in the DOS.
In addition, in the $s_{++}$-wave state, the peak in the DOS is robust against impurities.
Therefore, the absence of the coherence peak cannot be explained by the $s_{++}$-wave state and we can expect that the gap has the $s_{+-}$-wave symmetry.
However, in order to determine the pairing symmetry more clearly, measuring both the density and spin responses at the {\it single} wave vector $\bf{Q}$ (nesting vector) is essential.

\section{\label{5}Discussion}

In Sec. \ref{3}, we mentioned the resonance peak in inelastic neutron scattering measurements. There we pointed out that within the one-loop approximation or within the RPA used in Refs. [\citen{Korshunov}-\citen{Maier2}], the peak structure in inelastic neutron spectrum is obtained for $\omega\approx|\Delta^{(0)}_{\nu}|+|\Delta^{(0)}_{\nu'}|$ in the $s_{+-}$-wave state, not in the $s_{++}$-wave state. On the other hand, in a recent study, Onari, Kontani and Sato showed that a prominent hump structure in inelastic neutron scattering spectrum appears at $\omega\approx|\Delta^{(0)}_{\rm{min}}|+|\Delta^{(0)}_{\rm{max}}|$ in the $s_{++}$-wave state by considering the effect of the quasiparticle damping\cite{Sato}. Thus, the peak for $\omega\approx|\Delta^{(0)}_{\nu}|+|\Delta^{(0)}_{\nu'}|$ in inelastic neutron scattering experiment may not always the direct evidence for the $s_{+-}$-wave state. A more detailed inelastic neutron scattering measurement at $\omega\approx|\Delta^{(0)}_{\nu}|+|\Delta^{(0)}_{\nu'}|$ is needed for the resolution of this problem.

\section{\label{6}Conclusion}

For iron pnictides, we have theoretically proposed an experimental method which distinguishes between the $s_{+-}$- and $s_{++}$-wave states.
In each of these two states, we have analyzed both the density and spin response functions by using the two-band BCS model within the one-loop-approximation.
As a consequence, we found that in the $s_{+-}$-wave ($s_{++}$-wave) states the density (spin) response function has a peak just below $T_{c}$,
and the spin (density) response function decreases monotonically as temperature decreases.
Thus, the temperature dependence of the response functions in the $s_{+-}$-wave state is qualitatively different from that in the $s_{++}$-wave state.
We discussed that these results can be explained by the coherence effect which depends on the relative phase between the two superconducting gaps.
Moreover, from the analysis of effect of impurities, the additional information for the determination of the pairing symmetry has been obtained.
Our suggestion is that the $s_{+-}$- or $s_{++}$-wave state could be identified by measuring the temperature dependence of both the two kinds of response functions (two kinds of structure factors in actual experiments) at the nesting vector $\bf{Q}$ in neutron scattering measurements.
The theoretically expected behaviors of both the responses can be categorized into the following three classes:
(i) The density response has a peak just below $T_{c}$, and the spin response monotonously decreases as temperature decreases. In this case, we can determine that the gap of this material has the $s_{+-}$-wave symmetry without impurities (i.e., with negligible impurities).
(ii) The density response exhibits a monotonic decrease as temperature decreases, and the spin response shows a peak just below $T_{c}$. In this case, we can determine that the gap of this material has the $s_{++}$-wave symmetry either with or without impurities.
(iii) Both the density and spin responses show a monotonic decrease as temperature decreases. In this case, we can determine that the gap of this material has the $s_{+-}$-wave symmetry with impurities.

\section*{Acknowledgments}
We would like to thank Y. Akamine, Y. Yunomae, D. Inotani, D. Yamamoto, N. Yokoshi, and S. Tsuchiya for valuable comments and discussions.
We would also like to acknowledge T. Shigenari and Y. Tsunoda for helpful discussions on the experimental aspects.


\begin{thebibliography}{9}
\bibitem{remark1} In inelastic neutron scattering measurements, measuring $\omega$ dependence near $\omega=|\Delta^{(0)}_{\nu}|+|\Delta^{(0)}_{\nu'}|$ of the spin response for $T{\ll}T_{c}$ is of necessity.
\bibitem{remark2} In contrast to the case of the resonance peak ($\omega\approx|\Delta^{(0)}_{\nu}|+|\Delta^{(0)}_{\nu'}|$), measuring the temperature dependence of both the density and spin responses with $\omega{\ll}|\Delta^{(0)}_{\nu}|+|\Delta^{(0)}_{\nu'}|$ is of necessity.
\end{thebibliography}
\end{document}